\documentclass[3p,sort&compress,twocolumn]{elsarticle}

\usepackage{amssymb}
\usepackage{amsmath}
\usepackage{dsfont}

\journal{J.~Comput.~Phys.}

\newcommand{\bo}{\ensuremath{\boldsymbol{B}_0}}
\newcommand{\abs}[1]{\ensuremath{\left| #1\right|}}

\newcommand{\be}{\begin{equation}}
\newcommand{\ee}{\end{equation}}
\newcommand{\bs}{\begin{subequations}}
\newcommand{\es}{\end{subequations}}

\newcommand{\ez}{\ensuremath{\hat{\boldsymbol{e}}_z}}

\newcommand{\pa}{\ensuremath{_\parallel}}

\newcommand{\De}{\ensuremath{\varDelta}}

\newcommand{\La}{\ensuremath{\varLambda}}

\newcommand{\usum}{\ensuremath{\sum_{n=-\infty}^\infty}}

\newcommand{\f}[1]{\ensuremath{\boldsymbol{#1}}}
\newcommand{\m}[1]{\ensuremath{\left\langle #1\right\rangle}}

\newcommand{\df}{\ensuremath{\mathrm{d}}}

\begin{document}

\begin{frontmatter}
\title{On Simplified Numerical Turbulence Models in Test-particle Simulations}

\author[1]{R.\,C. Tautz\corref{cor1}}
\ead{rct@gmx.eu}

\cortext[cor1]{Corresponding author}
\address[1]{Zentrum f\"ur Astronomie und Astrophysik, Technische Universit\"at Berlin, Hardenbergstra\ss{}e 36, D-10623 Berlin, Germany}


\begin{keyword}
turbulence \sep plasma physics \sep Monte-Carlo simulation \sep diffusion
\end{keyword}

\end{frontmatter}

\section{Introduction}

Understanding the scattering of cosmic rays in the interplanetary and interstellar plasmas is a problem of central importance in astrophysics. The diffusion processes of charged particles in the directions, parallel and perpendicular, to an ordered magnetic field (e.\,g., the magnetic field of the Sun) can be described by the diffusion tensor, whose components can be (1) calculated using analytical transport theories; (2) extracted from numerical test-particle simulations; and (3) obtained from heliospheric observations. Understanding of such observations is a key subject of space physics \cite{bie94:pal,dro00:rig}.

An important way to test analytical theories \cite{jok66:qlt,tau06:sta} is numerical Monte-Carlo simulations \cite{gia94:mul,mic96:alf,gia99:sim,mic01:sim,tau10:pad} that operate under the same restrictions imposed on analytical calculations---stationary or static turbulence, prescribed turbulence geometry and power spectrum, no back-reaction of the particles on the turbulence field. Hence, such simulations are called ``test particle'' simulations although, using the same approach, other effects such as magnetic field line diffusion can be investigated.

In this Note, the special case of \emph{isotropic} magnetostatic turbulence will be investigated, which is an important test case for both numerical and analytical approaches \cite{fis74:iso,bie88:iso,tau06:sta,sha09:hil}. It will be shown what the basic turbulence properties are and how magnetic turbulence is usually generated in numerical simulations (Sec.~\ref{iso}). Some problems will be discussed that are inherent in the basic formulation of numerical turbulence, because not all physical requirement such as vanishing magnetic divergence and isotropy in position and wavenumber space can be fulfilled at the same time. Finally, several simulation results will be compared and discussed (Sec.~\ref{res}).

\section{Isotropic Turbulence}\label{iso}

Isotropic turbulent magnetic fields can be thought of as a superposition of plane waves with random phase angles and random orientations. In the limit of an infinite number of plane waves, the resulting turbulence is homogeneous and isotropic \cite{bat82:tur}. There are a number of both analytical and numerical constraints and also pitfalls. Consider each in turn.

\subsection{Analytical constraints}

In general, homogeneous turbulence \cite{bat82:tur} is described using a stochastic approach that is based on a two-point, two-time correlation tensor
\be
\bigl\langle B_l(\f x,t)\,B_m^\star(\f x',t')\bigr\rangle=\mathsf R_{lm}(\f x,\f x',t,t'),
\ee
where $B_{l,m}$ refers to the turbulent magnetic field components with $l,m\in\{x,y,z\}$.

To account for a power spectrum that is (at least partially) known in wavenumber space \cite{kol41:tur,kol91:tur}, a Fourier transform is applied, where the assumption of homogeneity leads to a delta function $\delta(\f k-\f k')$ \cite{bat82:tur,rs:rays,tau10:rag}. With the additional assumption of a time-independent turbulence field (i.\,e., magnetostatic turbulence), the result reads
\be
\bigl\langle\hat B_l(\f k)\,\hat B_m^\star(\f k')\bigr\rangle=\delta(\f k-\f k')\,\mathsf P_{lm}(\f k).
\ee

For homogeneous and isotropic turbulence, the correlation tensor has the form \cite{bat82:tur,rs:rays,sha09:nli}
\be\label{eq:iso_tens}
\mathsf P_{lm}(\f k)=\frac{G(k)}{8\pi k^2}\left(\delta_{lm}+\frac{k_lk_m}{k^2}+i\sigma\epsilon_{lmn}\,\frac{k_n}{k}\right)
\ee
with $\sigma(k)\in[-1,1]$ the magnetic helicity (usually assumed to be zero, with some noticeable exceptions) and $\epsilon_{lmn}$ the Levi-Civit\`a tensor.

The normalization of the correlation tensor $\mathsf P_{lm}$ is given through the condition \cite{rs:rays}
\be\label{eq:bnorm}
B^2=B_x^2+B_y^2+B_z^2=\int\df^3k\sum_{i=1}^3\mathsf P_{ii}(\f k),
\ee
where $B$ corresponds to the (average) turbulent magnetic field strength.

Consider now the three constraints for the isotropic turbulent magnetic field.

\paragraph{Wave vectors} Isotropic turbulence means that each orientation of the wave vector has equal probability. Consider the Fourier transform (for illustration purposes in two dimensions) of an isotropic function
\begin{align}
\hat F(\f k)&=\int\df^2r\;F(r)e^{i\f k\cdot\f r}\nonumber\\
&=\int_0^\infty\df r\;rF(r)\int_0^{2\pi}\df\phi\;e^{ikr\cos(\psi-\phi)},
\end{align}
with polar coordinates $\f k=(k\cos\psi,k\sin\psi)$ and $\f r=(r\cos\phi,r\sin\phi)$. Then \cite{gr:int}
\begin{align}
\int_0^{2\pi}\df\phi\;e^{ikr\cos(\psi-\phi)}&=\usum i^nJ_n(kr)\int_0^{2\pi}\df\phi\;e^{in(\psi-\phi)}\nonumber\\
&=2\pi\,J_0(kr)
\end{align}
so that the result does not depend on the orientation of the wave vector as described through the angle $\psi$. Here, $J_n$ denotes the Bessel function of the first kind of order $n$.

For the three-dimensional case, a similar calculation is slightly more involved. However, one can always transform to a new coordinate system $(r,\alpha,\beta)$ where $\alpha$ is defined through $\f k\cdot\f r=kr\cos\alpha$. Then $\beta$ is unused in the Fourier integral so that the two-dimensional case is recovered.

\paragraph{Field strength} According to Eq.~\eqref{eq:bnorm}, the mean value of each individual field component is determined through\footnote{Strictly speaking, $k=0$ must be excluded because it represents a uniform magnetic field.}
\be
B_i^2=\int\df^3k\;P_{ii}(\f k).
\ee
Due to the fact that the isotropic turbulence tensor from Eq.~\eqref{eq:iso_tens} does not distinguish, for example, the $z$ from the $x,y$ directions it is immediately clear that, on average, $\langle B_x^2\rangle=\langle B_y^2\rangle=\langle B_z^2\rangle$. Any isotropic turbulence generator must therefore fulfill the constraint of equal amplitude field components in all three spatial directions.

\paragraph{Divergence}

Every magnetic field must obey Maxwell's equation of vanishing divergence, i.\,e., $\nabla\cdot\f B=0$ corresponding to the property that no magnetic monopoles exist (although, in the realm of quantum effects, the subject remains under active investigation \cite{gib11:mon}).

\subsection{Numerical turbulence generation}\label{sim}

Following the ideas of \cite{gia94:mul,mic96:alf,gia99:sim,mic01:sim}, the turbulence in the \textsc{Padian} code \cite{tau10:pad} is generated via a summation over $N$ plane wave modes as
\be\label{eq:dB}
\f B(x,y,z)=\text{Re}\sum_{n=1}^N\hat{\f\xi}_nA(k_n)e^{i\left(k_nz'+\beta_n\right)},
\ee
where $\beta_n$ is the phase angle of the plane waves. The vector $\hat{\f\xi}_n$ denotes the amplitude direction of each wave and is defined as
\be
\hat{\f\xi}_n=\cos(\alpha_n)\hat{\f e}_{x',n}+i\sin(\alpha_n)\hat{\f e}_{y',n}.
\ee
where $\alpha_n$ is the polarization angle.

The unit vectors $\hat{\f e}_{x',n}$ and $\hat{\f e}_{y',n}$ are given by the first and second lines, respectively, of a three-dimensional rotation matrix
\be
\La_{lm}=
\begin{pmatrix}
\;\cos\theta\cos\phi & \cos\theta\sin\phi & -\sin\theta\;\\
\;-\sin\phi & \cos\phi & 0\;\\
\;\sin\theta\cos\phi & \sin\theta\sin\phi & \cos\theta\;
\end{pmatrix}.
\ee

Physically, one can think of the angle $\alpha$ describing wave types that vary between fast-mode waves ($\alpha=0$) and Alfv\'en waves ($\alpha=\pi/2$), but since no time-dependence is considered, the analogy is limited.

The direction of propagation of the plane waves, i.\,e., the $z'$ direction, results from $z'=x\,\La_{31}+y\,\La_{32}+z\,\La_{33}$. Because the $z'$ direction is always perpendicular to $\hat{\f\xi}_n$, one immediately has $\f k_n\cdot\hat{\f\xi}_n=0$ for every mode $n$, which corresponds to $\nabla\cdot\f B=0$, thus ensuring that the turbulent magnetic field is divergence free.

For each summand $n$, all angles $\theta$, $\phi$, $\alpha$, and $\beta$ are randomly generated\footnote{Note that not the angle $\theta$ but instead its cosine, $\eta=\cos\theta$, is uniformly distributed. This ensures that the density of wave directions is equal for all solid angles $\df\phi\,\df\eta$.} but are then kept fixed. Thus, the same $\f B$ results for the same set of coordinates $(x,y,z)$, corresponding to what is called a ``turbulence realization''.

By constraining the angles $\theta$ and $\alpha$ \cite{gia99:sim,tau10:pad}, non-isotropic turbulence geometries such as slab and 2D can be obtained. To fulfill the additional constraint that $\f B\perp\ez$ \cite{gra96:sca,bie96:two}, the polarization angle has to be set to $\alpha=\pi/2$.

The amplitude function $A(k_n)$ is defined through
\be
A^2(k_n)=G(k_n)\De k_n\left(\sum_{\nu=1}^NG(k_\nu)\De k_\nu\right)^{\!-1},
\ee
where, for example, the turbulence spectrum $G(k_n)$ is of the form \cite{sha09:flr}
\be\label{eq:spect}
G(k_n)=\frac{k_n^{\;q}}{\left(1+k_n^{\;2}\right)^{(s+q)/2}},
\ee
where $q$ and $s$ are the energy range and inertial range spectral indices, respectively. A logarithmic spacing of the wavenumbers is commonly used so that $\De k_n/k_n$ is constant.

\subsection{Normalization of the turbulence}

The turbulent field as generated through Eq.~\eqref{eq:dB} should be normalized to unity.\footnote{Technically, such is due to the requirement that a unit magnetic field vector $\hat{\f e}_B$ is generated, which, in the equation of motion, is ``manually'' scaled with the factor $B/B_0$ to yield the requested turbulence strength relative to the mean magnetic field strength, $B_0$ \cite{tau10:pad}.} Therefore, it is required that, on average,
\be
B^2=\m{B_x^2+B_y^2+B_z^2}=1.
\ee
If such were not the case, the resulting transport parameters would be falsified, i.\,e., would be too large (small) if the magnetic field strength were to be smaller (greater) than unity.

However, on implementing Eq.~\eqref{eq:dB} as is, one finds that: (i) the average strength of the turbulent fields is considerably smaller than unity; (ii) the $B_z$ component is (also on average) smaller than the other two components, which contradicts the requirement of isotropy. Both of these drawbacks are due to the fact that
\begin{align}
\m{\xi_x^2}&=\bigl\langle\abs{\cos\alpha\cos\theta\cos\phi-i\sin\alpha\sin\phi}^2\bigr\rangle=3/8\nonumber\\
\m{\xi_y^2}&=\bigl\langle\abs{\cos\alpha\cos\theta\sin\phi+i\sin\alpha\cos\phi}^2\bigr\rangle=3/8\nonumber\\
\m{\xi_z^2}&=\bigl\langle\left(-\cos\alpha\sin\theta\right)^2\bigr\rangle=1/4.
\end{align}
If the turbulent magnetic field components are divided by the mean values of the unit vector components, i.\,e., by $\sqrt{3/8}$ and by $1/2$, respectively, then all components of the turbulent magnetic field have equal means and the total magnetic field strength is approximately unity, as required.

\subsection{Fulfilling physical constraints}\label{constr}

One is faced with a choice because one has three options:
\begin{itemize}
\item using the original turbulence generation mechanism; but the original form for the turbulence is not in agreement with equal mean values for the three magnetic field components, as required for isotropy;
\item normalizing the turbulence as described above; but the normalization factors are not compatible with the requirement that the divergence of the turbulent magnetic field be zero (note that one still has $\langle\nabla\cdot\f B\rangle=0$ on average);
\item renormalizing the turbulent wave vector via $k_{x,y}\to k_{x,y}\sqrt{3/8}$ and $k_z\to k_z/2$ so that $\f k_n\cdot\hat{\f\xi}_n=0\;\forall n$ is restored; but then the wave vector is not isotropic any more.
\end{itemize}

\begin{figure}[tb]
\centering
\includegraphics[width=85mm]{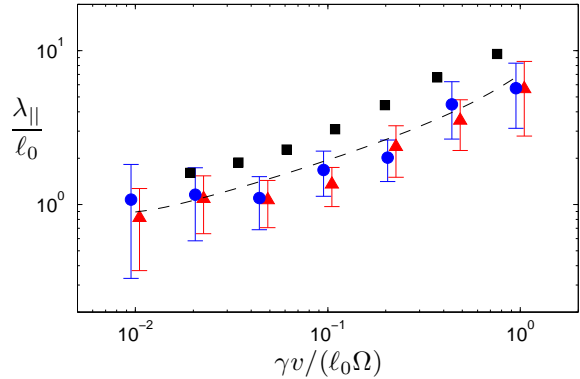}
\caption{(Color online) The parallel scattering mean free path as a function of the particle speed. The errorbars with red triangles and blue dots show the simulation results from the \textsc{Padian} code \cite{tau10:pad} with and without renormalized wave vectors, respectively. Previous simulation results \cite{gia99:sim} do not use any normalization of the turbulent magnetic field (black squares). The black dashed line shows an analytical result \cite{tau08:soq}.}
\label{ab:mfp}
\end{figure}

The important point to note is that, using the approach described by Eq.~\eqref{eq:dB}, not all three requirements can be fulfilled at the same time. Such is comparable, e.\,g., to various formulations of smoothed particle hydrodynamics \cite{vau08:sph,spr10:sph}, where conditions such as $\nabla\cdot\f B=0$ or the conservation of mass, energy, or angular momentum are frequently violated. It is therefore left to the resulting transport parameters to decide which option gives the best results.

\section{Results and Conclusion}\label{res}

In Fig.~\ref{ab:mfp}, the parallel mean free path is shown as resulting from two test-particle simulations in isotropic turbulence with a nominal turbulence strength $B=1$. Furthermore, the turbulent wave vector was scaled so that either $\f k_n\cdot\hat{\f\xi}_n=0$ or the orientation of $\f k$ is isotropic (see Sec.~\ref{constr}). However, such has only marginal influence on the resulting transport parameters compared to the estimated errors.

In contrast, comparison of the classic results by Giacalone \& Jokipii \cite{gia99:sim}, where no turbulence renormalization had been done, with the \textsc{Padian} results shows systematical deviation, as clearly exhibited by Fig.~\ref{ab:mfp}. Therefore, the main deviation results from the fact that, in the \textsc{Padian} code, the turbulent magnetic field components have been normalized so that $B=1$ on average.

Moreover, analytical results that have been derived using second-order quasi-linear theory \cite{sha05:soq,tau08:soq} agree better with test-particle simulations in a turbulent field with the correct turbulence strength. Such can be understood from the well-known fact that, as a rough estimate from classic quasi-linear theory \cite{jok66:qlt}, one has $\lambda\pa\propto(B/B_0)^{-2}$, thus underlining the important influence of the magnetic field strength on transport parameters. Here, $\bo=B_0\ez$ denotes the mean magnetic field, which is usually assumed to be homogeneous.

To conclude, using the conventional approach of superposing plane waves, it is not possible to create a strictly \emph{isotropic} turbulent magnetic field structure that obeys all physical constraints, which are (i) equal mean of all magnetic field components; (ii) isotropy of the wave vectors; and (iii) vanishing divergence of the magnetic field. Such magnetic fields are widely implemented in test-particle Monte-Carlo simulations, which are used to obtain (i) scattering mean free paths of charged particles; (ii) field line diffusion coefficients.

While the turbulent magnetic field strength plays an important role for the results, such does not seem to be the case for a non-zero magnetic field divergence and/or the isotropy of the wave vectors. Future work should explore the possibility of a turbulence approach that is sufficiently simple but is fully compatible with all physical boundary conditions.

\section*{Acknowledgments}

The author thanks Andreas Shalchi, Ian Lerche, and Timo Laitinen for valuable comments.




\end{document}